# Design of double-negative metamaterials using single-layer plasmonic structures

**Ling Lin and Richard J. Blaikie***

MacDiarmid Institute for Advanced Materials and Nanotechnology,

Department of Electrical and Computer Engineering, University of Canterbury,

Christchurch 8020, New Zealand

*Corresponding author, e-mail: r.blaikie@elec.canterbury.ac.nz; fax: +64(3)3642-761

**Abstract:** We propose a design for optical-frequency double-negative metamaterials (DMNs) in which the unit cell is formed by a single-layer plasmonic inclusion. By introducing some properly arranged stepwise modulations onto the inclusions, the effective permittivity and permeability of the material may become simultaneously negative due to the excitation of higher-order multipole plamonic (electrostatic) resonances. The approach opens a possibility to construct low-loss optical DNMs with great engineering simplicity.

# 1. Introduction

Double-negative metamaterials (DNMs) refer to artificially patterned metal-dielectric structures having simultaneously negative effective permittivity $\varepsilon_{eff}$ and effective permeability $\mu_{eff}$. Owing to their unique physical properties and the fascinating applications [1, 2, 3] that can not be accessed in naturally-occurring materials, DNMs have attracted much attention within the scientific and engineering communities since the results of Pendry, Smith, and coworkers [4, 5]. However, compared with engineering DNMs at larger wavelengths such as at microwave region, obtaining optical DNMs (particularly optical magnetism) is still a difficult task. As the electromagnetic (EM) properties of the metals at microwave region and at optical region are rather different, there exists a physical limitation on achieving magnetism at high frequencies by simply downsizing the microwave DNM elements such as split-ring resonators [6]. Nevertheless, the constituent units of optical DNMs generally involve nano-scaled features, which also bring in fabrication challenges.

In the optical region metallic nanostructures are often referred to as plasmonic structures due to their ability to support surface plasmon excitations. This feature has been extensively used in designing optical metamaterials. A common approach is based on the excitation of the symmetric and asymmetric plasmon resonances in the paired elements, such as arrays of metallic nano-rod pairs [7, 8] and paired perforated metal films [9,10,11], to obtain electrical and magnetic response of the structure, hence the DNMs. However, the antiresonant behaviour of the material parameters originating from the periodicity of the structure [12] increases the difficulty in overlapping the frequency bands of magnetic and electric resonances in those structures. While some authors suggested that such a limitation could be overcome by combining continuous metal film(s) with arrays of paired elements [13, 14], the complexity involved in such structures implies that the realization of the designs is a rather challenging task. On the other hand, Shvets *et al.* [15, 16] have proposed that a metal-dielectric composite consisting of closely-packed plasmonic inclusions can also generate strong magnetic activity leading to a modified magnetic permittivity $\mu_{eff} \neq 1$ when the frequency of operation approaches the higher-multipole plasmonic (electrostatic) frequency of the inclusions. In this manuscript we present a new design of a single-layer optical DNM based on this approach. The building element of structure is a subwavelength metallic inclusion with stepwise modulations which act as strong scatters to excite higher-order multipole electrostatic resonance. Double-negative behaviour of the structure occurs in the spectral range where these higher-order modes dominate over the dipole mode. This design may open up a simple alternative for experimental implementation of a DNM since it involves only one patterned metallic layer.

## 2. Structure Design

The structure consists of a single layer of 2D periodic metallic inclusions in the *x-y* plane. The inclusions have a finite thickness *h*, which is much smaller ($h < 100$ nm) than the wavelength of the incident light λ, along the *z*-direction. A schematic of the unit cell configuration (top view) is shown in Fig. 1. The basic element of the unit cell is a rectangular prism whose dimensions in the *x-y* plane are labeled as $w_o$ and $l_o$; some stepwise modulations, whose dimensions are indicated as $w_1$, $w_2$ and $w_3$ along *x*-direction and $l_1$, $l_2$ and $l_3$ along *y*-direction, are then introduced on both sides of this elementary rectangular unit to function as strong scatterers to produce higher-multipole plasmonic resonances for achieving a magnetic response. The characteristics (spectral position and strength) of the resonances are mainly governed by the shape of the inclusions due to its electrostatic nature, which offers great flexibility in tuning the effective parameters of the structure. While we use a three-step-modulation design to demonstrate the concept in this manuscript, the number of modulation steps can be introduced to the system are not restricted to the three steps shown in Fig. 1.

## 3. Simulation Results

A rigorous coupled-wave analysis (RCWA) [17] based commercial grating analysis tool, G-Solver [18] was used to calculate the complex transmission and reflection coefficients, $t$ and $r$, of the structures. The incident light has electric component $E_x$ and magnetic component $H_y$; the light propagates along the $z$ direction. Both the incident and outgoing media are air. The material for the inclusions was assumed to be Ag and its complex dielectric constants were taken from Ref. [19]. After obtaining $t$ and $r$, the complex refractive index $n$ and impedance $Z$ of the structure were calculated from [20]

$$\cos(nk_o h) = \frac{1 - r^2 + t^2}{2t} \quad (1)$$

and

$$Z = \sqrt{\frac{(1+r)^2 - t^2}{(1-r)^2 - t^2}}, \quad (2)$$

where $k_o$ is the propagation number of the incident light in free space and $h$ is the thickness of the media; and the effective $\varepsilon$ and $\mu$ were then retrieved using

$$\varepsilon = n/Z \quad \text{and} \quad \mu = nZ. \quad (3)$$

The EM properties of the structures were first analyzed by examining the effects of different geometric parameters of the inclusion, such as the thickness of the structure $h$, the number of modulation steps and the dimensions of each step, on the effective $n$, $\varepsilon$ and $\mu$ of the system; the results were then used to optimize the arrangement of the system to obtain a double-negative feature and the highest possible value of the figure-of-merit $-\text{Re}\{n\}/\text{Im}\{n\}$.

Fig. 2 show the simulation results for one of the optimized designs whose geometric parameters are listed in Table 1. The results reveal two key characteristics of the structure: the lowered electron plasma frequency $\omega_p$ of the metal ($\varepsilon = 0$ at ~ 820 nm wavelength); and simultaneously negative $\varepsilon$ and $\mu$ at the wavelengths around 1100 nm to 1130 nm. Fig. 2a reveals that around the wavelength corresponding to $\omega_p$, the structure has very high transmission efficiency; when moving towards the longer wavelength side of the spectrum, it shows the metal-like behaviour (highly reflective). A shoulder in the transmission curve occurs at the wavelengths between 1100 nm to 1150 nm and the transmission falls rapidly beyond this region, which coincides with the double-negative characteristics (hence the negative refractive index) of the structure in this spectral region: the transmitted light exhibits a negative phase velocity when propagating through a negative refractive medium, and the destructive interference between the incident and the transmitted light will cause a sharp decrease in the transmission around the negative index region. The phenomenon has also been observed experimentally [10]. Meanwhile, the real part of the impedance has a peak whereas the imaginary part shows a modulation (Fig. 2b) in this region. As a result of better impedance matching between the structure and the surrounding medium (air), a sharp dip in the reflection appears. Figure 2c shows that the real part of $\varepsilon$ exhibits some oscillations at 1000 nm to 1300 nm, which indicates the excitation of certain plasmon resonances within the structure; nevertheless, the effective $\mu$ of the structure show a strong modulation in this spectral region. As shown in Fig. 2d, the real part of $\mu$ ranges from −2.3 to 3.6. Furthermore, similar as reported in other designs of DNMs [12], in the spectral range where the magnetic resonance occurs, the imaginary part of $\mu$ also becomes negative over a narrow spectral region.

Owing to the existence of double-negative behaviour, the designed structure has very low losses in negative index region. As displayed in Fig.3, the real part of $n$ has a minimum value of $-4.6$, and the highest value of $-\text{Re}\{n\}/\text{Im}\{n\}$ reaches 4.2 at $\lambda = 1150$ nm, which is rather attractive for future physics experiments and application development.

To verify the origin of the double-negative behaviour observed in the single-layer plasmonic structure, we analyzed its near-field distribution using a finite element method (FEM) based commercial software package (COMSOL Multiphysics). Fig. 4 shows the electric and magnetic field distributions across the center of the structure ($z = ½\,h$ plane) at the magnetic resonance ($\lambda = 1126$ nm, $\text{Re}\{\mu\} = -2.3$). Clearly, the $E_x$ component of the electric field and $H_y$ component of the magnetic field arise from the light scattered by the stepwise modulations presented on the inclusion. The field distributions of $E_x$ and $H_y$ both show quadrupole resonance profiles, which suggests that the double-negative behaviour of the structure was produced by the higher-order multipole electrostatic resonance.

We would also like to point out that in RCWA method used to perform the spectral analysis is based on the (finite) Fourier series expansion of the EM fields and material permittivity, which are solutions to Maxwells equations, inside the corrugated region. For highly conducting periodic inclusions it generally requires a high number of Fourier components to obtain accurate results [21]. However, for the calculations of 2D metallic gratings, G-Solver appears to be highly unstable as the number of orders increases. Therefore, the number of orders retained in the calculation here were restricted to 2 in order to obtain meaningful results. Consequently, the amounts of uncertainty existing in these results remain unclear at this stage and more reliable simulation codes need be developed in the near future. Nevertheless, we note that the physical effects predicted from RCWA simulations have been confirmed independently using FEM techniques (which are unwieldy for performing the spectral calculations directly), so we have some confidence that the behaviour of these new structures is at least qualitatively correct.

## 4. Discussion and Conclusion

We have shown that it is possible to realize an optical DNM using a single-layer plasmonic structure. The simultaneously negative $\varepsilon_{\text{eff}}$ and $\mu_{\text{eff}}$ of the structure is a result of the excitation of higher-order multipole electrostatic resonances initiated by the strong scatters presented in the stepped subwavelength metallic inclusions. The approach offers great degrees of freedom in controlling of the EM properties of the structure. As these resonances are electrostatic in nature, their properties can be tailored by varying the shape of the inclusions. This approach could open up a new means of realizing low-loss optical DNMs with great engineering simplicity.

**Table Captions**

Table 1. Configurations of the unit cell of the structure used in this study.

**Table 1**

| $P$ (nm) | $h$ (nm) | $w_0$ | $l_0$ | $w_1$ | $l_1$ | $w_2$ | $l_2$ | $w_3$ | $l_3$ |
|---|---|---|---|---|---|---|---|---|---|
| 340 | 80 | 0.9P | 0.34P | 0.8P | 0.06P | 0.4P | 0.07P | 0.3P | 0.1P |

**Figure Captions**

Figure 1. A schematic diagram of the unit cell of a single-layer DNM.

Figure 2. (Color online) Simulation results for the device with geometrical parameters listed in Tabble 1 of: (a) transmittance and reflectance; (b) impedance; (c) effective permittivity; and (d) effective permeability.

Figure 3. (Color online) Retrieved refractive index $n$ of the designed structure. The insert shows the values of $-\text{Re}\{n\}/\text{Im}\{n\}$ around the negative-index region.

Figure 4. (Color online) Field distributions across the center of the unit cell ($z = \frac{1}{2} h$ plane) at magnetic resonance ($\lambda = 1126$ nm).

**Figure 1**

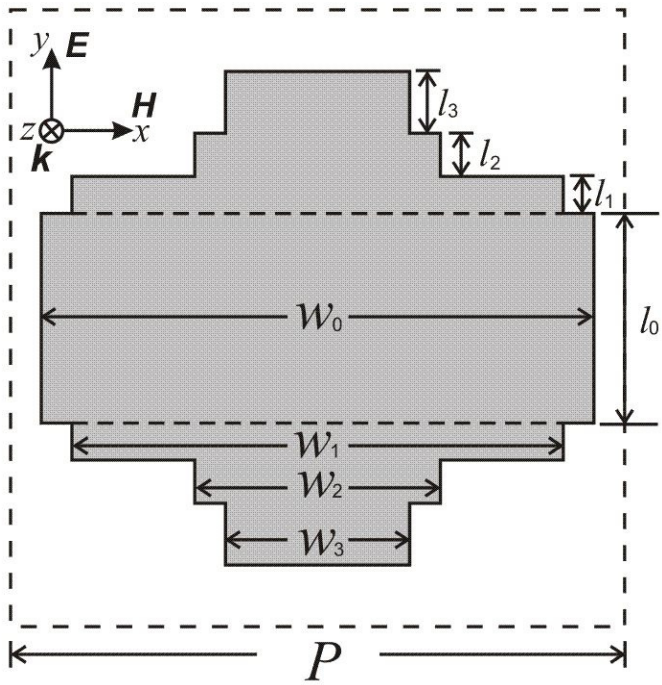

**Figure 2**

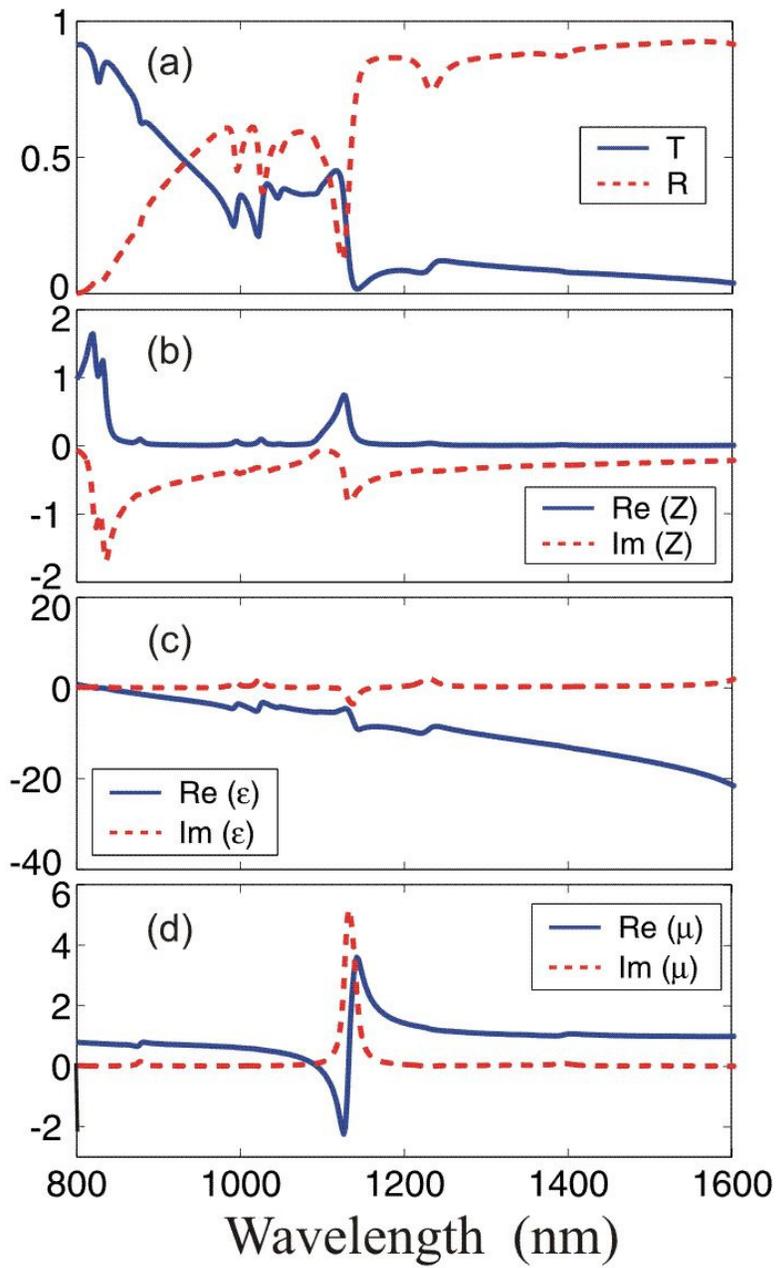

**Figure 3**

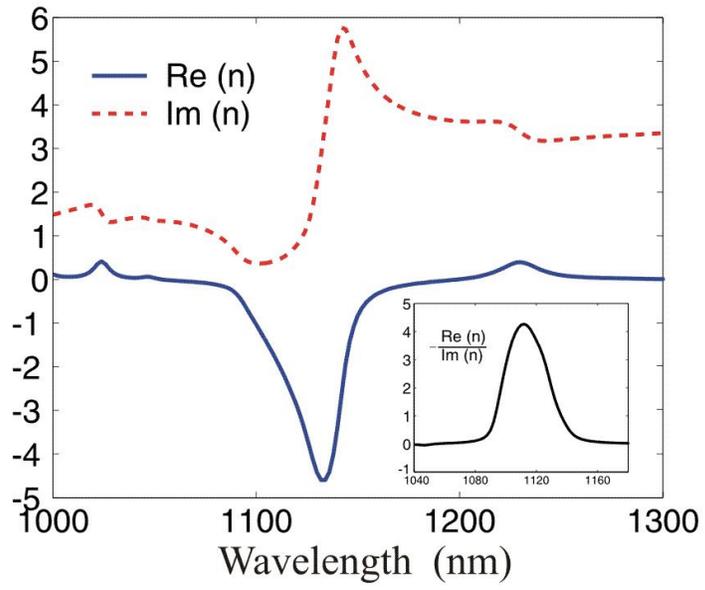

**Figure 4**

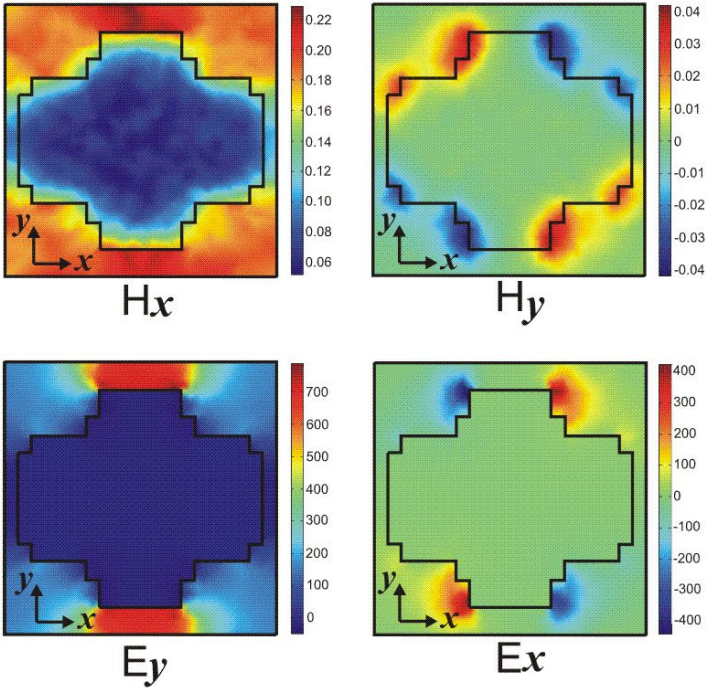